\documentclass[a4paper,11pt]{article}
\usepackage[utf8]{inputenc}
\usepackage{amsmath}
\usepackage{graphicx}
\usepackage{caption}
\usepackage{subcaption}
\usepackage{amssymb}
\usepackage{color}
\usepackage{cancel}
\usepackage{framed}
\usepackage{comment}
\usepackage[margin=1in]{geometry}
\usepackage{mathtools}

\newcommand{\nn}{\nonumber}

\def\s{\sigma}   
\def\z{\zeta}

\newcommand{\Acal}{{\mathcal A}} \newcommand{\Vcal}{{\mathcal V}}

\newcommand*{\affaddr}[1]{#1} 
\newcommand*{\affmark}[1][*]{\textsuperscript{#1}}

\def\Cincy{Department of Physics, University of Cincinnati, Cincinnati, Ohio 45221, USA}
\def\Fermilab{Fermi National Accelerator Laboratory, P.O. Box 500, Batavia, IL 60510, USA}

\begin{document}
\date{}
\title{\Large\bfseries QCD Axion Star Collapse with the Chiral Potential}

%

\author{%
Joshua Eby\affmark[*]\affmark[$\dag$], 
Madelyn Leembruggen\affmark[*], 
Peter Suranyi\affmark[*], and 
L.C.R. Wijewardhana\affmark[*] \vspace{.3cm} \\
{\it\affaddr{\affmark[*]\Cincy}}\\
{\it\affaddr{\affmark[$\dag$]\Fermilab}}\\
}

\begin{titlepage}
\maketitle

\begin{abstract}
In a previous work, we analyzed collapsing axion stars using the low-energy instanton potential, showing that the total energy is always bounded and that collapsing axion stars do not form black holes. In this paper, we provide a proof that the conclusions are unchanged when using instead the more general chiral potential for QCD axions.
\end{abstract}
\end{titlepage}

\section{Introduction}
Axion stars \cite{Tkachev,KolbTkachev,HoganRees,BB,ChavanisMR}, bound states of condensed axion particles \cite{PQ1,PQ2,Weinberg,Wilczek,DFS,Zhitnitsky,Kim,Shifman}, have recently received increased interest in the literature \cite{ESVW,Guth,Braaten}. Such states can have important effects in the context of cosmology and astrophysics, as well as fundamental particle physics \cite{Khlopov1,Khlopov2,Khlopov3,Khlopov4,Khlopov5,Preskill,Sikivie1,Davidson,DF,Holman,Sikivie2}. QCD axions, which are connected with the strong nuclear force, were originally proposed to solve the Strong $CP$ problem; such theories have a well-defined parameter space with a single free parameter, the axion mass. The masses of dilute QCD axion stars have an upper bound such that they are much lighter than an ordinary star \cite{ChavanisMR,ESVW,Guth}, though alternative axion theories can exceed this bound.

The possibility of a dense state for axion stars was first developed in \cite{Braaten}. If they exist, dense states can lead to novel astrophysical consequences, including new sources of photon emission through, e.g., accretion of hydrogen into dense configurations \cite{HAS}. The existence of these states depends on both attractive and repulsive interaction terms present in the axion potential. The well-studied dilute state, which is a metastable minimum of the energy, exists only for particle number $N<N_c$, a critical value. However, at any particle number $N$ in the condensate, the global minimum of the energy is at a radius many orders of magnitude smaller than that of the dilute state (though still larger than the corresponding Schwarzschild radius).

Lately, a significant amount of work has been done analyzing the stability properties of dilute axion stars, both in the context of gravitational collapse and decay through relativistic axion emission \cite{Tkachev4to2,Lifetime,ChavanisCollapse,ELSW,Tkachev2016,Marsh,Braaten2}. In a previous work \cite{ELSW}, we analyzed the collapse of an axion star using a variational method for finding energetically stable bound states. This analysis built on the formalism of \cite{ChavanisCollapse} (itself built on the previous work of \cite{ChavanisMR,PethikSmith,Perez}), who argued that boson stars with attractive self-interactions collapse to black holes. Our work \cite{ELSW}, which included both attractive and repulsive terms in the interaction potential for the axion field $\Acal$ \cite{nonchiral,Vecchia},\footnote{$m$ and $f$ are the mass and decay constant, respectively, of the underlying axion field theory. For QCD axions, typical values are $m = 10^{-5}$ eV and $f = 6\times10^{11}$ GeV.}
\begin{equation} \label{Vinstanton}
 V_I(\Acal) = m^2\,f^2\Big[1 - \cos\Big(\frac{\Acal}{f}\Big)\Big],
\end{equation}
concluded that axion stars do not collapse to form black holes. This was due to the existence of a dense state which is larger than the corresponding Schwarzschild radius. Further, number changing interactions inside the axion star \cite{Lifetime,Braaten2} may cause the collapsing axion star to emit a significant fraction of its energy in the form of relativistic free axions \cite{ELSW,Tkachev2016,Marsh}.

Our analysis in \cite{ELSW} made use of a common approximate form of the axion potential, sometimes called the ``instanton potential'', eq. (\ref{Vinstanton}) \cite{nonchiral,Vecchia}. It is known, however, that the system is more precisely described by the so-called ``chiral potential'' \cite{Vecchia,Cortona}
\begin{equation} \label{Vchiral}
 V_C(\Acal) = m^2\,f^2\,\frac{1+z}{z}\Big[1+z-
	  \sqrt{1+z^2+2\,z\,\cos\Big(\frac{\Acal}{f}\Big)}\Big].
\end{equation}
The latter form takes into account the effect of nonzero light quark masses through the parameter $z=m_u/m_d$, where $m_u$ ($m_d$) is the mass of the up (down) quark. Note that $\lim_{z\rightarrow0}V_C = V_I$. While $V_I$ is more tractable and lends itself more easily to an analytic expansion, it is possible that our conclusions (the boundedness of the energy, the existence of a dense global energy minimum) would have been different if we had used the more precise form $V_C$. We thus revisit the analysis with the improved potential here.

There is a second potentially limiting assumption in the analysis of \cite{ELSW}. We used a variational method to minimize the energy, which gives good agreement with other, more precise analyses \cite{BB,ESVW,Braaten,ChavanisNumeric} in analyzing bound states of axion stars. This method requires the input of some ansatz for the wavefunction $\psi$ of the axion star, defined by the non-relativistic expansion \cite{Guth,Nambu}
\begin{equation} \label{NRexp}
 \Acal(r) = \frac{1}{\sqrt{2\,m}}\Big[e^{-i\,m\,t}\,\psi(r) + e^{i\,m\,t}\,\psi^*(r)\Big].
\end{equation}
In that work, we analyzed in detail a Gaussian ansatz,
\begin{equation} \label{Gaussian}
 \psi(r) = \frac{\sqrt{N}}{\pi^{3/4}\s^{3/2}}e^{-r^2/2\s^2},
\end{equation}
and also presented the analogous result for a $\cos^2$ ansatz,
\begin{equation} \label{cos2}
    \psi(r) = 
  \begin{dcases}
    \sqrt{\frac{4\,\pi\,N}{(2\pi^2-15)R^3}}\cos^2\Big(\frac{\pi\,r}{2\,R}\Big),& 		r\leq R\\
    0,              & r>R
  \end{dcases}
\end{equation}
We presented both to provide evidence that our results were not an artifact of some particular choice of ansatz. Indeed, these two general classes, those functions which extend to $+\infty$ but decrease monotonically (as does the Gaussian) and those which have compact support (as does $\cos^2$), are those well-behaved enough to be considered reasonable choices for a ground-state axion star wavefunction. Nonetheless, it could be asked whether anything changes upon assuming only a general, well-behaved ansatz in one of these two classes. We thus revisit this question here as well.

This paper is organized as follows. We review the variational method in Section \ref{VarSec}, and use it to expand the chiral potential in a tractable form in Section \ref{ChiralSec}. We examine the boundedness of the energy in Section \ref{BoundSec}, and conclude in Section \ref{ConcSec}.

\section{Review of the Variational Method} \label{VarSec}
The total axion star energy, in the non-relativistic and weak-binding limits, can be written as \cite{ELSW}
\begin{equation} \label{Hamiltonian}
 E  = \int d^3r\,\Big[\frac{|\nabla\psi|^2}{2m} + W(\psi) 
		  + \frac{1}{2}V_{grav}\,|\psi|^2\Big]
\end{equation}
where $W(\psi)$ represents the effective axion self-interaction potential (in this case, either the non-relativistic limit of eq. (\ref{Vinstanton}) or eq. (\ref{Vchiral})), and
\[
 V_{grav}(|\psi|^2) = -G\,m^2\int 
	\frac{\psi^*(x^\prime)\psi(x^\prime)}{|x^\prime-x|}d^3x^\prime.
\]
The wavefunction $\psi$ in the variational method is approximated by some tractable ansatz. Typically the wavefunction will be monotonically decreasing, and can either extend to $+\infty$ or have compact support on a finite radius $R$. 

We will use the generic scaling of macroscopic parameters derived in \cite{ESVW}
\begin{equation} \label{scaling}
 R = \frac{1}{m}\,\frac{\rho }{\sqrt{\delta}} \qquad N = \frac{f^2}{m^2}\,\frac{n }{\sqrt{\delta}},
\end{equation}
where $\delta \equiv f^2/M_P^2 \approx 10^{-14}$ for QCD parameters, and the Planck mass is $M_P = 1.22\times10^{19}$ GeV. Then a generic ansatz can be written as
\begin{equation}\label{ansatz}
\psi(r)= \sqrt{m}\,f\,\zeta\, F\left(\frac{r}{R}\right)\equiv \sqrt{m}\,f\,\zeta\,F(\xi),
\end{equation}
which, without loss of generality, can be chosen such that $F(0)=1$. The dimensionless normalization constant $\zeta$ is fixed by the requirement that $\int \psi^*\psi \,d^3r = N$; that is,
\begin{equation} \label{zetaeq}
 \zeta = \sqrt{\frac{\delta\,n}{C_2\,\rho^3}}
\end{equation}

Then the total energy can be written as
\begin{equation}\label{binding}
 \frac{E(\rho)}{m\,N} = \delta\left(\frac{D_2}{2\,C_2}\frac{1}{\rho^2}-\frac{B_4}{2\,C_2{}^2}\,\frac{n}{\rho}-\frac{n}{\rho^3}\,v\right),
 \end{equation}
where 
\begin{equation}\label{vform}
v = 4\,\pi\,\frac{\rho^6}{n^2\,\delta^2}\int d\xi\,
      \xi^2 \,\frac{W(\psi)}{m^2\,f^2}
\end{equation}
is a rescaled self-interaction energy, and where we defined the functions
\begin{align}
B_4 &= (4\,\pi)^2\int d\xi\,\xi \,F(\xi)^2\int_0^\xi d\eta\, \eta^2\,F(\eta)^2 \\
C_k &= 4\,\pi\int d\xi\,\xi^2 \,F(\xi)^k \\
D_2 &= 4\,\pi\int d\xi\,\xi^2 \,F'(\xi)^2.
\end{align}
For a given ansatz, the functions $B_4$, $C_k$, and $D_2$ are numerical constants.

\subsection{The Instanton Potential}
In \cite{ELSW}, we used the instanton potential of eq. (\ref{Vinstanton}) to investigate the bound states in the axion star energy. In that case the rescaled functional $v$ was calculated in closed form, both as an integral and also as an infinite series:
\begin{align} \label{vformInstanton}
v_I &= 4\,\pi\,\frac{\rho^6}{n^2\,\delta^2}\int   d\xi\,\xi^2
	    \left[1-J_0\left(\sqrt{2}\,\zeta\,F(\xi)\right)
			-\frac{\zeta^2}{2}F(\xi)^2\right] \\
 &= \sum_{k=0}^\infty \left(-\frac{1}{2\,C_2}\right)^{k+2}
	    \left(\frac{n\,\delta}{\rho^3}\right)^k\frac{C_{2\,k+4}}{[(k+2)!]^2}.
\end{align}
We used these simplified expressions to show that the total axion star energy was always bounded from below, and that the global minimum of the energy in eq. (\ref{binding}) was at a radius much larger than the corresponding Schwarzschild radius.

A similar analysis, using the chiral potential of eq. (\ref{Vchiral}), is somewhat more challenging. However, making use of some convenient analytic properties of the self-interaction energy makes the problem tractable. We present such an analysis in the next section.

\section{The Chiral Potential} \label{ChiralSec}
The potential we wish to consider has the form \cite{Vecchia,Cortona}
\begin{equation} \label{VcNorm}
  \frac{V_C(\Acal)}{m^2\,f^2} \equiv \Vcal = \frac{1+z}{z}\Big[1+z-
	  \sqrt{1+z^2+2\,z\,\cos\Big(\frac{\Acal}{f}\Big)}\Big].
\end{equation}
Using the particle data group values of
\begin{align*}
m_u=2.15 \pm 0.15 \text{ MeV}\\
m_d=4.70 \pm 0.20 \text{ MeV}
\end{align*}
we find that
\begin{equation}
z=\frac{m_u}{m_d}\simeq 0.457.
\end{equation}
To recover the form of the instanton potential in eq. (\ref{Vinstanton}), one merely needs to take the $z\to0$ limit, which gives $\Vcal = 1-\cos(\Acal\,/\,f)$.

We will follow the procedure of \cite{ESVW} and write $\Acal = \Acal^+ + \Acal^-$, where $\Acal^+$ ($\Acal^-$) creates (annihilates) one axion in the ground state wavefunction. Expanding eq. (\ref{VcNorm}) in a power series of the cosine, we have
\begin{equation}\label{nseries}
\Vcal = \frac{1+z}{z}\left[1+z -\sqrt{1+z^2}\sum_{j=0}^\infty
      \binom{\frac{1}{2}}{j}\left(\frac{2\,z}{1+z^2}\right)^j
      \left[\cos\left(\frac{\Acal^+ + \Acal^-}{f}\right)\right]^j.\right]
\end{equation}
Now noting that in leading order of $N$, $\Acal^+$ and $\Acal^-$ commute,
\begin{equation}
\cos\left(\frac{\Acal^+ + \Acal^-}{f}\right)
      =\frac{1}{2}\left(e^{i\,\Acal^+/f}e^{i\,\Acal^-/f}+e^{-i\,\Acal^+/f}
	      e^{-i\,\Acal^-/f}\right).
\end{equation}
Expand the $j	$th power of the cosine,
\begin{align} \label{nthpower}
 \left[\cos\left(\frac{\Acal^+ + \Acal^-}{f}\right)\right]^j
    &= \frac{1}{2^j}\sum_{k=0}^j\binom{j}{k}
	  e^{i\,(j-2\,k) \Acal^+/f}e^{i\,(j-2\,k) \Acal^-/f} \nn \\
  &\rightarrow \frac{1}{2^j}\sum_{k=0}^j\binom{j}{k}
	\sum_{m=0}^\infty(j-2\,k)^{2\,m}
	(-1)^m\frac{1}{(m!)^2}\left(\frac{\Acal^+\Acal^-}{f^2}\right)^m \nn \\
  &=\frac{1}{2^j}\sum_{k=0}^j\binom{j}{k}J_0\left[(j-2\,k)\,
	\left(\frac{2\sqrt{\Acal^+\Acal^-}}{f}\right)\right].
\end{align}
where in the second step, we expanded the exponentials and kept only those terms with equal numbers of $\Acal^+$ and $\Acal^-$. Inserting this expression into eq. (\ref{nseries}) gives the self-interaction energy density as follows:
\begin{equation}\label{Vsum}
\Vcal = \frac{1+z}{z}\left\{1+z -\sqrt{1+z^2}\sum_{j=0}^\infty
      \binom{\frac{1}{2}}{j}\left(\frac{z}{1+z^2}\right)^j
      \sum_{k=0}^j\binom{j}{k}J_0\left[(j-2\,k)\,
      \left(\frac{2\sqrt{\Acal^+\Acal^-}}{f}\right)\right]\right\}
\end{equation}

Using the correspondence with the non-relativistic Gross-Pit\"aevskii approach~\cite{ELSW}, as defined by the expansion in eq. (\ref{NRexp}),
\begin{equation}\label{GPcomp}
\frac{2\sqrt{\Acal^+\Acal^-}}{f} = \sqrt{\frac{2}{m}}\frac{\sqrt{\psi^*(r)\,\psi(r)}}{f}
	= \sqrt{2}\,\zeta\,\Big |F\left(\frac{r}{R}\right)\Big |,
\end{equation}
we can obtain the self-interaction potential in terms of the rescaled wavefunction $F(\xi)$, as defined in eq. (\ref{ansatz}). The range of $\xi$ is either finite and then we can choose $\xi_{\rm max}=1$, or it is infinite, in which case we choose $F(\xi)= O(\xi^{-2})$, as $\xi\to\infty$.

\subsection{Integral Representation}
Our aim in this section is to perform the summations over $j$ and $k$ in eq. (\ref{Vsum}) at the price of introducing integrals, since the latter is easier to estimate. Using the ansatz of eq. (\ref{ansatz}), the rescaled self-interaction energy, defined in eq. (\ref{vform}), is
\begin{align} \label{vC}
 v_C =&4\,\pi\,\frac{\rho^6}{n^2\,\delta^2} \int d\xi\,\xi^2 \,
       \frac{1+z}{z}\sqrt{1+z^2}\sum_{j=0}^\infty
      \binom{\frac{1}{2}}{j}\left(\frac{z}{1+z^2}\right)^j \nn \\
	 &\times  \sum_{k=0}^j\binom{j}{k}
	  \Big\{1 - J_0\left[(j-2k)\sqrt{2}\,\zeta\, F(\xi)\right]  -\frac{(j-2k)^2\,\zeta^2}{2}\,F(\xi)^2\Big\},
\end{align}
so that the expression in the energy functional evaluates to
\begin{align}\label{VSI}
\frac{E_{SI}(\rho)}{m\,N} 
	&= -\frac{n\,\delta}{\rho^3}\,v_C \nonumber \\
	&= -\frac{1+z}{z}\sqrt{1+z^2}\sum_{j=0}^\infty\binom{\frac{1}{2}}{j}\left(\frac{z}{1+z^2}\right)^j\nonumber
	\\ &\times\sum_{k=0}^j\binom{j}{k}\,
	      \frac{4\,\pi}{\zeta^2\,C_2}\int_0^\infty d\xi\,\xi^2\,
	      \left\{1-J_0\left[(j-2\,k)\,\sqrt{2}\,
	      \zeta\,F(\xi)\right]-(j-2\,k)^2\frac{\zeta^2}{2}\,F(\xi)^2\right\}
\end{align}
First, we will perform the summation over $k$ by using the standard integral representation
\begin{equation}\label{intrep}
J_0[(j-2\,k)\,u]=\frac{1}{\pi}\int_0^\pi \,e^{i\,(j-2\,k)\,u\,\sin t}dt
\end{equation}
 for $J_0$.
Substituting (\ref{intrep}) into (\ref{VSI}) we can perform the summation over $k$, yielding
\begin{align}\label{VSI3}
\frac{E_{SI}}{m\,N} 
   &= -\frac{1+z}{z}\sqrt{1+z^2}\sum_{j=0}^\infty\binom{\frac{1}{2}}{j}
		  \left(\frac{2\,z}{1+z^2}\right)^j\nonumber\\
   &\times\frac{4\,\pi}{\zeta^2\,C_2}\int_0^\infty
	  \left[1-\frac{1}{\pi}\int_0^\pi 
	  [\cos(\sqrt{2}\,\zeta \,F(\xi)\,\sin t)]^j\,dt-j\,
	    \zeta^2\frac{F(\xi)^2}{2}\right] \xi^2\,d\xi.
\end{align}
Now the summation over $j$ can also be performed, to yield the following expression for the energy functional
\begin{align}\label{VSI4}
\frac{E_{SI}}{m\,N} &= -\frac{4\,\pi}{\zeta^2\,C_2}\int_0^\infty 
	I[\sqrt{2}\,\zeta\,F(\xi)]\, \xi^2\,d\xi+\frac{1}{2},
\end{align}
where
\begin{equation}
I(u)=\frac{1+z}{z\,\pi}\int_0^\pi \left[1+z-\sqrt{1+z^2+2 \,z\,\cos(u\,\sin t)}\right]dt.
\end{equation}

It is easy to see that the integrand is bounded for all $u$ and for all $t\in\{0,\pi\}$, and that it is monotonic, being minimized at $z\rightarrow 0$. But the $z\to 0$ limit of $I(u)$ is nothing but the rescaled instanton potential
\begin{equation}
\lim_{z\to0} I(u)= 1 - J_0(u).
\end{equation}
The boundedness of the instanton potential therefore implies the boundedness of the chiral potential. It thus suffices to show that the instanton self-interaction potential is bounded from below. We showed exactly this in the Appendix of \cite{ELSW}, but only for the Gaussian ansatz. For completeness, we prove the general case in the next section.

\section{Boundedness of the Energy} \label{BoundSec}
In \cite{ELSW}, the proof that the axion star energy had a dense global minimum depended on the boundedness of the axion self-interaction potential, which we were able to show in the specific case of a Gaussian wavefunction. This proof can be generalized to any wavefunction which falls into one of two categories: For $F(\xi)$ defined in eq. (\ref{ansatz}), either (a) $\xi$ has a  finite range, such that $F(\xi)=0$ at $\xi \geq 1$ (i.e. the axion star has compact support on radius $R$), or (b) $\xi$ has an infinite range but $F(\xi)\rightarrow 0$ as $\xi\rightarrow \infty$.\footnote{To our knowledge, all relevant ans\"atze used by other authors fall into one of these categories as well.} We also assume that $F(\xi)$ is monotonically decreasing, though even that condition could be relaxed. This could be relevant, for example, for rotating axion stars, whose wavefunctions have nodes at the origin.

In this section we prove, under general conditions on the ansatz, that the self-interaction term in the axion star energy is bounded, and the kinetic energy term $~ 1\,/\,\rho^2$ is dominant as $\rho\rightarrow 0$.

The general form of  ans{\"a}tze for the wave function, which we and other authors use is of the form in eq. (\ref{ansatz}). As shown in eq. (\ref{binding}), such an ansatz leads to the following equation for the self-interaction energy
\begin{equation}
  \frac{E_{SI}(\rho)}{m\,N}
	= -\frac{\delta\,n}{\rho^3}\,v_I
	= -\frac{4\,\pi}{\zeta^2\,C_2}\,\int d\xi\, \xi^2
	      \left\{1-J_0[\sqrt{2}\z\,F(\xi)]-\frac{\z^2}{2}F(\xi)^2\right\},
\end{equation}
  
 The $\rho\to0$ limit of the self-energy is given by the $\z\to\infty$ limit, as seen in eq. (\ref{zetaeq}). The the integral 
\begin{equation}
  K = \frac{1}{\z^2}\int d\xi\, \xi^2\left\{1-J_0[\sqrt{2}\z\,F(\xi)]
	- \frac{\z^2}{2}F(\xi)^2\right\}.
 \end{equation}
The last term of $K$
\begin{equation}
  -\frac{1}{\z^2}\int d\xi\, \xi^2\,\frac{\z^2}{2}F(\xi)^2
      = -\frac{C_2}{8\,\pi}
\end{equation}
is always finite, since we must have\footnote{For example, in \cite{ELSW}, we had $C_2=\pi^{3/2}$, evaluated using the Gaussian ansatz.}
\begin{equation}
 C_2=4\,\pi\int_0^\infty d\xi\,\xi^2\,F(\xi)^2< \infty,
 \end{equation}
convergent for a normalizable wavefunction. Consequently, we can restrict ourselves to investigate the $\z\to\infty$ behavior of the remainder of the integral
\begin{equation}\label{I}
 K^\prime = \frac{1}{\z^2}\int d\xi\, \xi^2
	  \left\{1-J_0[\sqrt{2}\,\z\,F(\xi)]\right\}.
\end{equation}
  
The multiplier $1-J_0[\sqrt{2}\,\z\,F(\xi)]$ of the integrand of $K$ is positive, and bounded by its value taken at $\sqrt{2}\,\z\,F(\xi)=j_{1,1}=3.83171$, where $j_{1,1}$ is the first zero of $J_1(x)$. Then we have the inequality
\begin{equation} \label{BesselBound}
 1-J_0[\sqrt{2}\,\z\,F(\xi)]
      \leq B
      \equiv 1-J_0(j_{1,1})
      = 1.40276.
\end{equation}

Let us consider ans\"atze in the first class (a).  In that case, using the bound $B$ in eq. (\ref{BesselBound}), we have 
\begin{equation}
 K^\prime = \frac{1}{\z^2}\int_0^1 d\xi\, \xi^2
	\left\{1-J_0[\sqrt{2}\,\z\,F(\xi)]\right\}
	\leq \frac{1}{\z^2}\int_0^1 d\xi\, \xi^2 B
    = \frac{B}{3\,\z^2}\sim \rho^3.
\end{equation}
Clearly, $K^\prime \to 0$ as $\rho\to0$.
  
Now consider ans\"atze in class (b). We break up  integral $K^\prime$ of eq. (\ref{I}) such that in $K_1^\prime$ the region of integration is  $0\leq \xi\leq \xi_1$  while in $K_2^\prime$  it is $\xi>\xi_1$. We fix $\xi_1$, such that $\sqrt{2}\,\z\,F(\xi_1)=\nu$, where  $\nu\ll1$, a constant.  Then using again the bound in eq. (\ref{BesselBound}), we certainly have
\begin{equation}\label{bound1}
  K_1^\prime \leq \frac{\xi_1{}^3\,B}{3\,\z^2}.
\end{equation}
 Then it is easy to see that there is an $a>0$ such that
\begin{equation}\label{Fbound}
  F(\xi)< \frac{a}{\xi^{3/2}}.
\end{equation}
Taken at $\xi=\xi_1$, eq. (\ref{Fbound}) can be inverted as
\begin{equation}
 \xi_1{}^3< \frac{a^2}{F(\xi_1)^2}=\frac{a^2\,\z^2}{\nu^2}.
\end{equation}
Substituting $\xi_1$ into eq. (\ref{bound1}) we obtain that $K_1^\prime < a^2\, B\,/\, 3\,\nu^2$, i.e. it has a finite limit as $\z\to \infty$.
 
Now consider integral 
\begin{equation}\label{I2}
 K_2^\prime = \frac{1}{\z^2}\int_{\xi_1}^\infty d\xi\, \xi^2\left\{1-J_0[\sqrt{2}\,\z\,F(\xi)]\right\}.
\end{equation}
As $\sqrt{2}\,\z\, F(\xi_1)= \nu \ll1$ and as $F(\xi)$ is monotonically decreasing as a function of $\xi$, the argument of the Bessel function in eq. (\ref{I2}), $\sqrt{2}\,\z\,F(\xi)\ll 1$ at all $\xi$. Then we can safely expand the Bessel function and keep terms only up to the second order, to get
\begin{equation}
  K_2^\prime \simeq \frac{1}{\z^2}\int_{\xi_1}^\infty d\xi\, \xi^2  \,
	\frac{\z^2}{2}\,F(\xi)^2< \frac{C_2}{8\,\pi},
\end{equation}
bounded.  Then $K^\prime=K_1+K_2$ is also bounded. We conclude that, for any generic, monotonically decreasing ansatz for the axion star wavefunction, the instanton potential self-energy is bounded below. Because the chiral potential, as a function of $z$, is bounded below by the instanton potential, we conclude that the chiral potential is also bounded.

\section{Conclusions} \label{ConcSec}
In this work, we have extended the analysis of \cite{ELSW} to two important cases, namely, to the more precise chiral potential for axions, and to a generic ansatz for the wavefunction in the variational method. In both of these cases, we have found that the general conclusions of \cite{ELSW} hold: collapsing axion stars do not form black holes, but rather they are stabilized in a dense state which is a global minimum of the axion star energy. 

The analysis in this work, and the majority of \cite{ELSW}, are performed in the non-relativistic limit; we have neglected relativistic effects which lead to stimulated emission of axions through number changing operators \cite{Lifetime,Tkachev2016,Marsh,BraatenRelativistic,Iball}. The possibility exists that these dense states are unstable to emission of relativistic axions; we will return to this topic in a future publication.

 \section*{Acknowledgements}
 We thank P. Argyres, P. Fox, R. Gass, R. Harnik, A. Kagan, and M. Ma for conversations. The work of J.E. was supported by the U.S. Department of Energy, Office of Science, Office of Workforce Development for Teachers and Scientists, Office of Science Graduate Student Research (SCGSR) program. The SCGSR program is administered by the Oak Ridge Institute for Science and Education for the DOE under contract number DE-SC0014664.


\end{document}